\documentclass[conference]{IEEEtran}
\IEEEoverridecommandlockouts

\usepackage{cite}
\usepackage{amsmath,amssymb,amsfonts}
\usepackage{algorithmic}
\usepackage{graphicx}
\usepackage{textcomp}
\usepackage{xcolor}
\def\BibTeX{{\rm B\kern-.05em{\sc i\kern-.025em b}\kern-.08em
    T\kern-.1667em\lower.7ex\hbox{E}\kern-.125emX}}
\begin{document}

\title{G-AMC: A Green Automatic Modulation Classification Method}

\author{\IEEEauthorblockN{Chee-An Yu}
\IEEEauthorblockA{\textit{University of Southern California}\\
Los Angeles, USA \\
cheeanyu@usc.edu}
\and
\IEEEauthorblockN{Young-Kai Chen}
\IEEEauthorblockA{\textit{Coherent Corp.} \\
New Providence, New Jersey, USA\\
Young-Kai.Chen@Coherent.com}
\and
\IEEEauthorblockN{C.-C. Jay Kuo}
\IEEEauthorblockA{\textit{University of Southern California}\\
Los Angeles, USA\\
jaykuo@usc.edu}

}

\maketitle

\begin{abstract}
In this work, we propose an efficient and transparent green learning pipeline to address the automatic modulation classification (AMC) problem. This pipeline aims to enable receivers to blindly identify the modulation modes of the incoming signals in a computationally efficient way with a small model size. Our method includes the following steps. First, the input signal is transformed into a precise representation through the sparse coding method. Second, various features are extracted from the sparse coding representation with the statistics from the input signal. Third, the classification subspace is hierarchically partitioned with a tree structure to achieve a lightweight model size with good prediction accuracy. The experimental results demonstrate the effectiveness and efficiency in classifying the modulated features and representation of received signals. Compared to lightweight deep learning models,  the number of model parameters is reduced by \textbf{41\%}, while the usage of Floating Point Operations (FLOPs) is only $\mathcal{O}(10^{-4})$ of the blind waveform recognition without pre-arranged knowledge of incoming waveforms. 
\end{abstract}


\section{Introduction}

\begin{figure}[htbp]
\centering
\includegraphics[width=0.9\columnwidth]{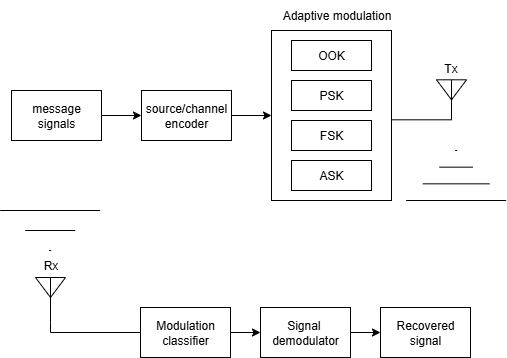}
\caption{Overview of the adaptive modulation framework.}
\label{overview of AMC}
\end{figure}

Automatic modulation classification (AMC), which aims to predict the modulation type of received signals, plays an essential role in enhancing the flexibility, adaptability, and intelligence of modern communication systems, particularly in cognitive radio, military, and dynamic spectrum environments. The overview of the adaptive modulation framework that incorporates an AMC module is illustrated in Fig.\ref{overview of AMC}\cite{huynh2021automatic}. AMC enables adaptive functionality in a receiver by automatically detecting modulation schemes to reconfigure the receiver to capture the incoming signal in real time. AMC enables blind recognition of signals without prior coordination or pre-trained metadata under non-cooperative scenarios. This would improve the efficiency of shared spectrum usage by identifying signals and modulations already in use to mitigate co-channel interferences. In addition, it can improve security and surveillance capability by classifying unknown or malicious signals\cite{zheng2025recent}. 
Traditionally, there are two categories of AMC algorithms: likelihood-based (LB) methods \cite{wei2000maximum,sills1999maximum} and feature-based (FB) methods\cite{wu2008novel,ramkumar2009automatic,zhu2014genetic, wang2010fast,wang2017graphic}. The likelihood method compares the received signal against known probabilistic models from a set of original raw data sets with known modulation types to determine the most likely modulation mode. Although FB can achieve optimal estimation in a Bayesian sense, it is sensitive to mismatches in model assumptions and has a high computational cost. In contrast, FB extracts features by calculating the statistical results in either the time domain or the transform domain from the original raw data. Due to its robustness and low computational cost, FB has become a more practical and favored method than LB methods. After extracting the discriminative features, the classifier can be applied to achieve AMC. Recently, the classification accuracy has improved significantly with the advancement of deep neural networks (DNNs), characterized by end-to-end training and superior contextual representation\cite{he2016deep}. In DNNs, both the feature extraction and the classifiers are integrated into a mysterious-box model architecture, introducing an inherent uncertainty and lack of interoperability in its decision-making and outcomes. Also, the large model sizes and computing complexity of DNNs may not be affordable on edge devices.

In this work, compared to other deep learning architectures for end-to-end training with backpropagation, we design a feedforward-modulated pipeline under a green learning framework that emphasizes model interpretability and computational sustainability\cite{kuo2023green}. In particular, we implement a learning-based method for the AMC problem with explainable features with a small model size and much reduced computational resources. The pipeline consists of three steps. First, the input signal is represented by a sparse coding method\cite{lee2006efficient} to form a suitable basis of different modulation patterns. Second, statistical features are extracted from the input signal and the sparse coding representation. They summarize the modulation patterns in different time ranges. Third, hierarchical tree-based classifiers are utilized to reduce redundant parameters and perform modulation classification. This modularized processing pipeline is mathematically transparent with intermediate deterministic results, which provides good explainability and analysis for security communication applications. In addition, it is computationally efficient and provides low power consumption, which is suitable for edge devices in practical applications. The green learning pipeline of this work is summarized in the following sections.

\section{Current methodologies}

This chapter reviews the two typical LB and FB methods and recent learning-based methods. The motivation and pros and cons of these methods will be illustrated.

\subsection{Likelihood-based method}

LB methods treat the AMC problem as a decision-theoretic framework \cite{ wei2000maximum,sills1999maximum}. In this framework, each modulation hypothesis is evaluated by its likelihood given the observed signal, and classification is achieved by maximizing likelihood. It is the optimal classifier in the Bayesian sense, but suffers from a high computational cost due to the evaluation of likelihoods for multiple candidate modulations\cite{ge2021accuracy}. In addition, it had to assume prior knowledge of parameters such as carrier frequency, phase, and symbol rate, making it less robust in practice. To improve the robustness of the LB method, \cite{abu2018automatic} proposed a hybrid scheme combining statistical moments. Incorporation of features based on invariant moments with FB showed effectiveness in mitigating sensitivity to unknown channel effects. Generally, FB methods have a solid grounding in statistical decision theory and provide understandable error bounds. However, practical use is limited due to robustness issues and computational burden. These limitations motivated the exploration of more flexible and practical FB methods. 

\subsection{Feature-based method}

Feature-based (FB) methods showed the effectiveness of mitigating sensitivity to unknowns. Instead of maximizing the likelihood of the waveform, FB methods extract discriminative signal measurements as characteristics and then apply a classifier to these characteristics to decide the modulation format. As a result, FB methods generally operate with lower complexity and less prior information than LB. The trade-off is that FB is typically suboptimal in the Bayesian sense\cite{ghasemzadeh2018performance} but can be much more practical for real-world use.

A rich variety of features has been proposed over the years. For example, high-order statistics\cite{wu2008novel}, cyclostationary features\cite{ramkumar2009automatic}, time-frequency distributions\cite{zhang2019automatic}, constellation diagrams\cite{wang2017graphic}, cumulative distribution\cite{zhu2014genetic, wang2010fast}, and time-frequency features\cite{zhang2019automatic} are commonly used in practice. These features are designed to capture invariant representations of nuisance parameters such as channel gain or phase offset from the received signal. Although FB methods are more robust and practical for the AMC problem, the process of feature design and decision-making rules heavily relies on human experts and manual tuning. Therefore, the labor-intensive feature engineering process motivates the development of learning features from data.

\subsection{Learning-based method}

With the advancement of machine learning, especially for deep learning, the accuracy of AMC improves dramatically. Unlike hand-crafted feature approaches, deep learning models (e.g., convolutional neural networks) can learn raw signal representations with the guidance from label information by the backpropagation algorithm. Deep learning models include CNNs, RNNs, and transformers, which process raw signal representations directly, surpassing traditional methods in accuracy under various SNR conditions \cite{huang2019automatic, hamidi2021mcformer, zheng2023mobilerat}.

\cite{chang2021multitask} proposed a CNN-RNN multitask learning architecture to fuse I/Q and amplitude-phase signal representations, improving classification accuracy. Transformers have also shown great promise. In \cite{hamidi2021mcformer}, MCFormer was introduced for its ability to capture long-term correlation in signals. In addition, residual networks\cite{peng2023automatic} demonstrated the effectiveness of masked modeling to enhance robustness in wireless communication scenarios. Moreover, attention mechanisms have been effectively utilized in frequency domain analysis. \cite{zhang2023frequency} introduced frequency learning attention networks, improving accuracy by adaptively focusing on informative frequency components. \cite{hao2023automatic} applied meta-learning to AMC, achieving improved adaptability and generalization with minimal labeled data. Unsupervised and semi-supervised approaches have also gained attention.\cite{ali2017unsupervised} explored unsupervised feature learning to reduce dependence on labeled datasets. Similarly, \cite{ali2021automatic} employed contrastive learning within a fully convolutional framework, achieving effective feature extraction and robust classification.

Despite these advancements, deep learning methods face challenges and concerns regarding computational efficiency, generalizability, and interpretability\cite{kim2021deep}. It is difficult to deploy high-complexity models on edge devices. In addition, the superior performance of deep learning models comes from heavy supervision of the label data. Recently, the models have been trained with synthetic data due to the high cost of real-labeled data. The domain shift between synthetic and real data should be overcome to achieve generalization. Furthermore, the black-box end-to-end training process makes the intermediate results meaningless. The decision process becomes unclear, and the features in the middle stage become untraceable.

\section{Green Learning Pipeline methodology}

\begin{figure}[htbp]  % or [b] for bottom placement
\centering
\includegraphics[width=1\columnwidth]{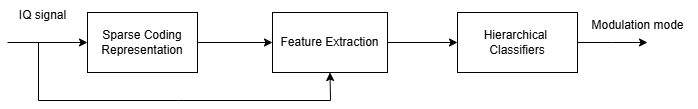}
\caption{Overview of green automatic modulation classification pipeline.}
\label{overview_pipeline}
\end{figure}

An overview of our green automatic modulation mode (G-AMC) classification pipeline is illustrated in Fig. \ref{overview_pipeline}. It consists of three modules: 1) sparse coding representation, 2) feature extraction, and 3) hierarchical classifiers. Module 1 is tailored to find the basis of different modulation patterns and separate them from the noisy signal. Module 2 integrates human design features and learning representations from sparse coding. Then, Module 3 is a hierarchy of small classifiers to achieve automatic modulation mode classification. 

\subsection{Sparse Coding Representation}

\begin{figure}[htbp]  % or [b] for bottom placement
\centering
\includegraphics[width=1\columnwidth]{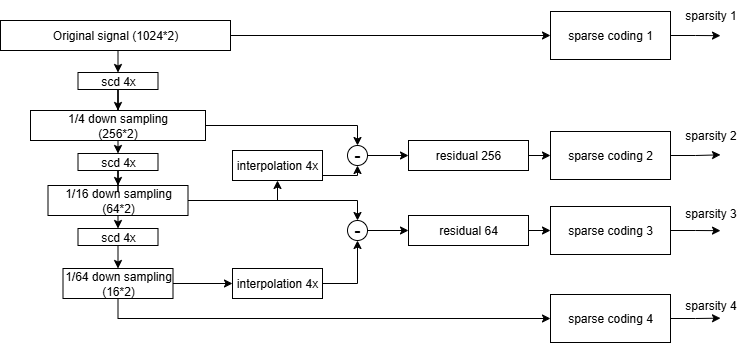}
\caption{Overview of the sparse coding representation.}
\label{spare_coding}
\end{figure}

The reasons for applying the sparse coding method to represent input signals are interpretability and denoising. The sparse dictionary learning algorithm aims to find a set of atoms and a projection on these atoms such that the input signals can be reconstructed. It can be formulated as the following optimization problem. Given dataset $X = [x_1, x_2,..., x_K ]\in R^d$. Find $D \in R^{d\times n}: D=[d_1, d_2, ..,d_n]$, $R\in R^n: R=[r_1, r_2,..., r_K]$ such that
\begin{multline}
\min_{\mathbf{D} \in \mathcal{C},\, \mathbf{r}_i \in \mathbb{R}^n}
\sum_{i=1}^{K} \left\| \mathbf{x}_i - \mathbf{D} \mathbf{r}_i \right\|_2^2 
+ \lambda \left\| \mathbf{r}_i \right\|_0, \\
\text{where } \mathcal{C} \equiv \left\{ \mathbf{D} \in \mathbb{R}^{d \times n} :
\left\| \mathbf{d}_i \right\|_2 \leq 1 \;\forall i = 1, \dots, n \right\}, \\
\lambda > 0
\end{multline}
Based on the assumption that noise signals require many atoms, sparse coding representations are applied to signal denoising\cite{oja1999image}. The sparsity representation forces signal preservation. Each atom in the learning dictionary represents a part of the signal from different modulation patterns. Different sparsity ($k$) controls the trade-off between the reconstruction error and the denoise effect. The higher the sparsity (lower $k$), the better the denoising effect. In this work, to consider different SNR conditions, we apply multiple values of the sparsity parameter $k$ to enrich the representation of the input signal. Furthermore, we design an inverse pyramid structure of multi-resolution signals to capture the global and transition modulation patterns. The overview of the sparse coding representation is illustrated in Fig. \ref{spare_coding}. The sparse coding 1 and sparse coding 4 are designed to represent the global patterns of the input signal, while the residual between different time resolutions extracts high frequency that represents the transition patterns of the input signal. To capture the transition modulation patterns, the residual signal is divided into several small windows, given the window size. The maximum pooling is applied for the local windows, which means the most significant projection value on the dictionary atoms. For example, the residual 64 with 64 samples in the time domain is cropped to 4 non-overlapped windows of size 16. Given a dictionary size of 64 atoms, each window can be represented by a 64-dimensional vector. Therefore, we can acquire 4 vectors in $R^{64}$. After applying maximum pooling in the window axis, we can acquire a vector in $R^{64}$ to represent the maximum response of each atom in each window. The transition patterns projected on the dictionary atoms are extracted by sparse coding 2 and sparse coding 3. The sparse coding representations consisting of global and local transition information are forwarded to the feature extraction module in the next step.

\subsection{Feature Extraction}\label{AA}

In the feature extraction module, we design several human-inspired features based on statistical results. First, the amplitude and phase histograms of the input signal are extracted as a statistical index to break the time-domain interference. The intuition behind the histogram feature design is that the modulation patterns are irrelevant to the symbol sequences. In particular, the feature representations of the modulation should be invariant to the order of the symbols. The amplitude and phase histograms capture the amplitude and phase distribution of the modulated signals. In addition, to tailor different SNR conditions, different bins of the histogram are applied to improve the robustness of the representation. For example, lower bin numbers present a robust representation in a low-SNR environment, while higher bin numbers describe detailed information of the signals in a high-SNR environment. Therefore, multi-histogram representations that apply different numbers of bins are designed as a feature. Furthermore, four non-linear derived features from histograms are designed to enrich the representations. Skewness, kurtosis, entropy, and variance are extracted from each histogram. These features can characterize amplitude-based modulation patterns. For example, kurtosis describes the peakedness of distributions. Therefore, the kurotis extracted from the AM amplitude histogram should have a lower value, while the PSK and FM modulations should have a larger value. Furthermore, we extract the high-order cumulants\cite{o2018over}, bispectrum\cite{dong2021radar}, cyclostationary\cite{camara2019automatic}, and wavelet packet energy\cite{liu2014multiwavelet} to improve the robustness in low SNR. Finally, sparse coding representations that contain both global and local information are concatenated with the above features as a feature vector in $R^{1730}$. Before feeding the feature vector to the hierarchical classifier, the discriminant feature test (DFT) \cite{yang2022supervised}, which is also a tool developed in the green learning pipeline, is applied to select discriminant features by removing redundant information. The summary of the extracted features and their dimensions is in TABLE \ref{tab:feature-dimensions}.

\begin{table}[h]
\caption{Feature dimension breakdown for green AMC pipeline}
\label{tab:feature-dimensions}
\centering
\begin{tabular}{|l|c|}
\hline
\textbf{Features} & \textbf{Dimension} \\
\hline
Amplitude and Phase Histogram (bins: 4, 8, 16, 32, 64) & 248 \\
Refinement Amplitude Histogram (bins: 128) & 130 \\
Phase Different Histogram (bins: 4, 8, 16, 32, 64) & 124 \\
FFT Histogram (bins: 4, 8, 16, 32, 64) & 248 \\
Statistics derived from Histograms & 40 \\
Circular Statistics & 3 \\
High Order Cumulants & 40 \\
Rotational moment (M=8,16,32) & 18 \\
Eigen Structure &  2 \\
Bispectrum  & 4 \\ 
Cyclostationary & 16 \\
Wavelet Energy & 16 \\
Amplitude CDF & 9 \\
Phase Different FFT & 128 \\
Sparse Coding Residual & 384 \\
Sparse Coding Global & 320 \\

\hline
\textbf{Total} & \textbf{1730} \\
\hline
\end{tabular}

\end{table}
\subsection{Hierarchical Classifiers}

\begin{figure}[htbp]  % or [b] for bottom placement
\centering
\includegraphics[width=1\columnwidth]{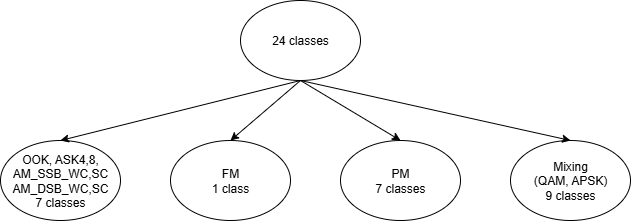}
\caption{Hierarchical Classifiers. The coarse classifier classifies all classes into four subclasses first. Then, each refinement classifier identifies the final classification result.}
\label{structure_classifier}
\end{figure}

Compared to designing a large classifier, we design several small classifiers in a hierarchical structure. The hierarchical structure achieves model simplification through the dedicated design of the sub-classes. The idea is that classifying easy samples/classes requires fewer parameters. That is, there are redundant parameters for the simple classes. Through the hierarchical tree-based classifier design, we can remove this redundancy and put more emphasis on the difficult classes. In the RadioML 2018.01A dataset\cite{o2018over}, we divide the original 24 classes into four sub-classes according to their modulation property. The first sub-class is relevant to amplitude modulations, the second is frequency modulation, the third is phase modulation, and the rest are the mixing of amplitude and phase modulations, like QAM and APSK. The structure of hierarchical classifiers is in Fig. \ref{structure_classifier}. According to this structure, the coarse classifier is responsible for classifying 4 classes to tell the types of modulations. Following the coarse classification, the refinement classifiers for each type are used to classify the final modulation result by separating modulation parameters. The first refinement classifier (R0) is for amplitude-related modulations. The second refinement classifier (R1) is for phase-related modulations, and the third refinement classifier (R2) is for the mixing modulations, such as QAM and APSK. The coarse classifier and refinement classifiers are all Extreme Gradient Boosting Trees (XGBoost) with the same parameters.

% TO DO:
% Tree-based algorithms: learning or rule-based...

\section{Experiment}

In this chapter, we verify each module in our G-AMC pipeline step by step. The experiments were run on the AMC EPYC 7543 32-core CPU with Ubuntu 22.04.   
\subsection{Dataset}
We use the RadioML 2018.01A dataset \cite{o2018over} that generates IQ signals using USRP B210 with SNR ranging from -10 dB to 20 dB with steps of 2 dB, where the Rayleigh channel model is applied. Common wireless communication conditions are modeled by random variables such as roll-off factor, carrier frequency offset, and symbol rate offset. The detailed range of randomness can be referenced in \cite{o2018over}. The dataset consists of 24 modulations as follows: 1) Digital modulation: OOK, ASK4, ASK8, BPSK, QPSK, PSK8, PSK16, PSK32, APSK16, APSK32, APSK64, APSK128, QAM16, QAM32, QAM64, QAM128, QAM256, GMSK.
2) Analog modulation: AM-SSB-WC, AM-SSB-SC, AM-DSB-WC, AM-DSB-SC, FM. Each modulation has 4096 frames for each SNR condition, and each frame has 1024 samples in the I and Q channels, respectively. Therefore, for each SNR, we have 98,304 frames and a total of 1,572,864 frames for SNR from -10 to 20 dB.

\subsection{Modulation Classification with Hierarchical Classifiers}

\begin{figure}[htbp]  % or [b] for bottom placement
\centering
\includegraphics[width=1\columnwidth]{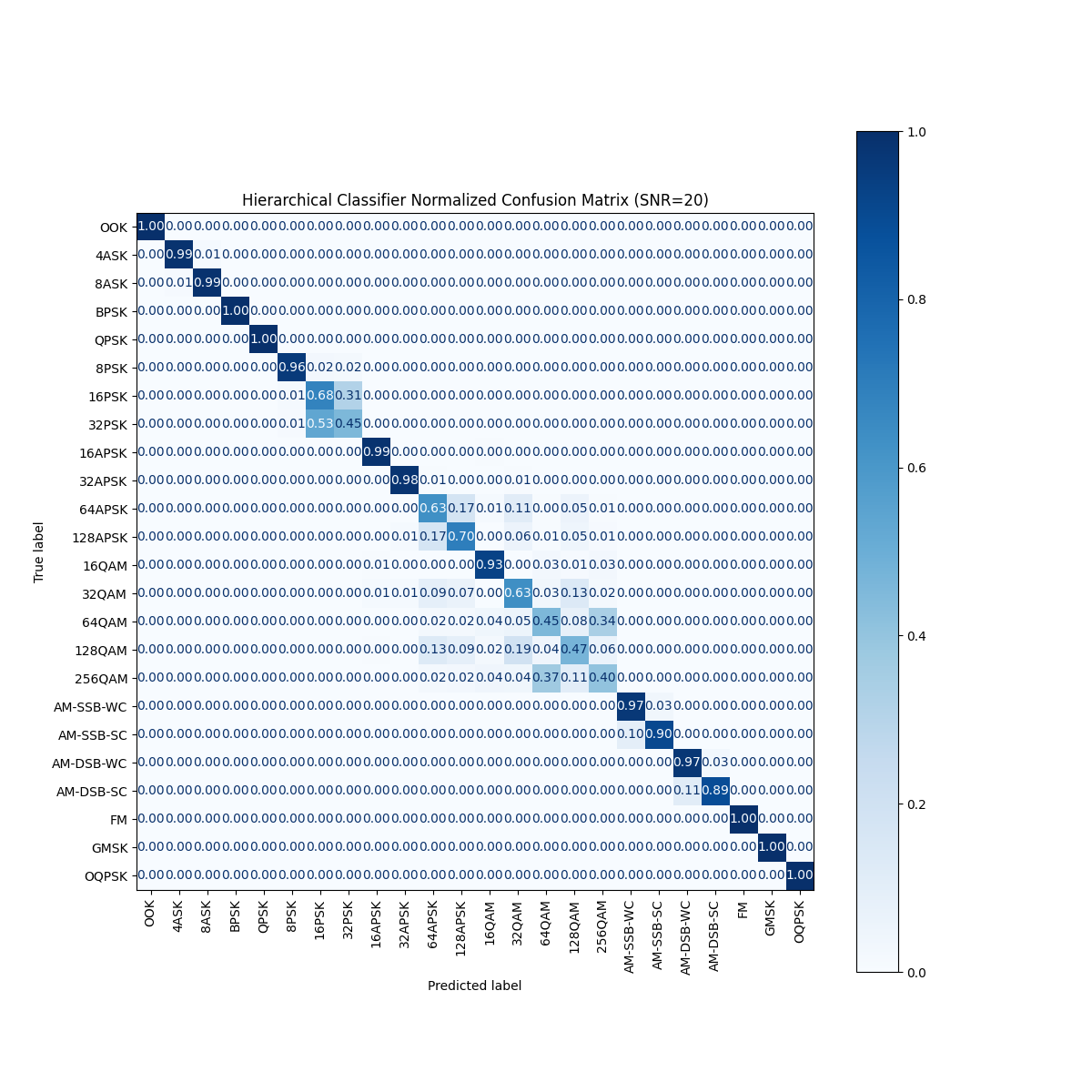}
\caption{Confusion matrix of Hierarchical classifiers under SNR=20.}
\label{confu_snr20}
\end{figure}

To train our hierarchical classifiers, we use 80\% of the data for the training process and 20 \% for the testing. The coarse XGBoost parameters of the coarse and refinement are shown in the Table \ref{tab:xgboost_params}. We report the accuracy of the coarse classifier and each refinement classifier individually, and their overall accuracy in the testing dataset in different SNR levels ranging from 20 to -10 dB in the Table \ref{tab:coarse_vs_overall}. The confusion matrix of the modulation classification result under SNR 20 is in Fig. \ref{confu_snr20}. We can see that the hierarchical classifier can generally predict the correct type of modulated signal from the coarse classifier. For refinement classifiers, the accuracy degrades at high-order modulation parameters, such as distinguishing 256QAM and 64QAM. One possible explanation is that there is a high overlap between 256QAM and 64QAM in the constellation graph. To classify them, we should sample the IQ components out of the overlapping region to characterize the high-order QAM. However, the short frame with 1024 samples does not have enough IQ samples from the non-overlapping region. 

\begin{table}[htbp]
\centering
\caption{XGBoost Hyperparameters}
\begin{tabular}{|l|c|}
\hline
\textbf{Parameter} & \textbf{Value} \\
\hline
Learning\_rate & 0.3 \\
Max\_depth     & 2 \\
N\_estimators  & 100 \\
objective      & mlogloss \\
\hline
\end{tabular}
\label{tab:xgboost_params}
\end{table}

\begin{table}[htbp]
\centering
\caption{Coarse classifier and overall accuracy across SNR levels}
\begin{tabular}{r|ccccc}
\textbf{SNR} & \textbf{Coarse \%} & \textbf{R0 \%} & \textbf{R1 \%} & \textbf{R2 \%} &\textbf{Overall \%} \\
\hline
-10 & 51.20 & 24.06 & 12.50 & 10.96 & 8.69 \\
-8  & 66.43 & 34.22 & 29.96 & 11.63 & 16.24 \\
-6  & 72.72 & 42.62 & 51.07 & 11.31 & 23.37 \\
-4  & 76.43 & 54.05 & 52.50 & 11.35 & 29.36 \\
-2  & 81.12 & 67.63 & 43.80 & 12.53 & 35.29 \\
 0  & 90.27 & 74.99 & 54.29 & 18.12 & 44.73 \\
 2  & 95.57 & 81.08 & 67.25 & 30.36 & 56.42 \\
 4  & 98.95 & 86.62 & 72.63 & 39.90 & 64.95 \\
 6  & 99.81 & 89.71 & 81.01 & 48.38 & 71.97 \\
 8  & 99.96 & 92.55 & 83.59 & 56.11 & 76.55 \\
10  & 99.97 & 94.38 & 84.75 & 61.65 & 79.51 \\
12  & 99.99 & 95.45 & 85.49 & 65.04 & 81.32\\
14  & 99.99 & 95.52 & 86.38 & 66.46 & 82.14 \\
16  & 99.98 & 95.67 & 87.15 & 67.03 & 82.61 \\
18  & 99.99 & 96.06 & 87.86 & 67.61 & 83.15 \\
20  & 99.98 & 95.85 & 86.97 & 68.77 & 83.27 \\

\end{tabular}
\label{tab:coarse_vs_overall}
\end{table}

\subsection{Model Size and Complexity Analysis}

We compared our model size in numbers of model parameters and complexity in terms of required Floating Point Operations Per Second (FLOPs) with other deep learning methods. G-AMC only required 27k model parameters. G-AMC uses only 19.2\% and 58\% of the numbers of parameters compared to those used in MCNet\cite{huynh2020mcnet} and light weight DLAMC \cite{kim2020lightweight}, respectively. In terms of FLOPs, G-AMC requires $\mathcal{O}(1.9 \times 10^{-3})$ FLOPs compared to CNN-AMC, which already represents one of the most lightweight deep learning approaches in terms of computation\cite{meng2018automatic}, as shown in TABLE \ref{tab:model_complexity}. It is also worth mentioning that G-AMC outperforms CNN-AMC for several SNR cases. These simulated results demonstrate the computing efficiency and sustainability of our G-AMC.

\begin{table}[h]
    \centering
    \caption{Model size and complexity analysis}
    \label{tab:model_complexity}
    \begin{tabular}{l|cc}

        \textbf{Model}              & \textbf{FLOPs}                     & \textbf{Model size} \\ 
        \hline
        VGG  \cite{o2018over}                       & 25{,}729{,}024 (3{,}176$\times$)   & 257 k (9.5$\times$)\\
        ResNet \cite{o2018over}                     & 10{,}688{,}512 (1{,}319$\times$)   & 236 k (8.74$\times$)\\
        CNN-AMC \cite{meng2018automatic}                     & 4{,}429{,}892 (516$\times$)        & 575 k (21.3$\times$) \\
        MCNet \cite{huynh2020mcnet}                          & 19{,}298{,}016 (2{,}382$\times$)   & 142 k (5.2$\times$) \\
        Lightweight DLAMC\cite{kim2020lightweight}         & 90{,}470{,}520 (11{,}169$\times$)  & 46 k (1.7$\times$) \\
        \hline
        G-AMC (Ours)                                        & 8{,}100 (1$\times$)                & 27 k (1$\times$) \\ 
    \end{tabular}
\end{table}

\begin{table*}[htbp]
\centering
\small
\caption{Classification accuracy (\%) across different SNR levels. The top three accuracy values are marked in bold, underline, and italic, respectively. }
\begin{tabular}{r|ccccccccccc}

\textbf{SNR} & \textbf{kNN} & \textbf{DT} & \textbf{SVM} & \shortstack{\textbf{ML-}\\\textbf{XGBoost}}\cite{o2018over} & \textbf{VGG}\cite{o2018over} & \textbf{ResNet}\cite{o2018over} & \shortstack{\textbf{CNN}\\\textbf{AMC}}\cite{abu2018automatic} & \shortstack{\textbf{MCNet}\cite{huynh2020mcnet}} & \shortstack{\textbf{Lightweight}\\\textbf{DLAMC}\cite{kim2020lightweight}} & \shortstack{\textbf{G-AMC}\\\textbf{(ours)}} \\

\hline
-10 & 4.39 & 7.90 & 8.60 & 7.74 & \textit{13.39} & \underline{14.23} & 10.47 & \textbf{15.19} & 10.97 & 8.69 \\
-8  & 4.44 & 8.69 & 9.52 & 8.06 & \underline{20.00} & \textit{19.86} & 16.23 & \textbf{22.32} & 17.39 & 16.24 \\
-6  & 6.42 & 10.78 & 12.90 & 9.03 & 24.03 & \underline{26.76} & 19.19 & \textbf{29.53} & \textit{26.66} & 23.37 \\
-4  & 11.96 & 15.84 & 18.15 & 12.74 & 30.16 & 30.70 & 26.02 & \underline{36.62} & \textbf{36.78} & 29.36 \\
-2  & 16.54 & 21.73 & 23.92 & 26.94 & \textit{37.26} & 36.20 & 30.59 & \underline{46.12} & \textbf{46.65} & 35.29 \\
0   & 21.19 & 26.49 & 28.00 & 38.71 & 44.84 & \textit{45.92} & 41.43 & \underline{53.51} & \textbf{56.64} & 44.73 \\
2   & 27.27 & 33.93 & 33.76 & 47.42 & 50.65 & 53.80 & 49.58 & \underline{62.38} & \textbf{65.32} & \textit{56.42} \\
4   & 32.32 & 41.82 & 40.87 & 51.94 & 54.84 & 61.69 & 54.51 & \underline{67.48} & \textbf{75.65} & \textit{64.95} \\
6   & 37.45 & 46.37 & 43.94 & 51.94 & 60.00 & 68.59 & 58.42 & \underline{74.45} & \textbf{84.01} & \textit{71.97} \\
8   & 40.20 & 48.39 & 46.97 & 58.23 & 62.42 & 70.00 & 60.49 & \underline{80.52} & \textbf{89.29} & \textit{76.55} \\
10  & 41.95 & 50.04 & 49.02 & 61.13 & 62.42 & 73.66 & 61.57 & \underline{86.06} & \textbf{91.48} & \textit{79.51} \\
12  & 42.46 & 50.95 & 50.72 & 62.42 & 62.90 & 73.94 & 62.07 & \underline{89.20} & \textbf{92.86} & \textit{81.32} \\
14  & 42.27 & 51.66 & 50.38 & 63.55 & 63.55 & 72.68 & 62.44 & \underline{89.86} & \textbf{94.21} & \textit{82.14} \\
16  & 43.46 & 51.71 & 51.05 & 62.42 & 64.19 & 73.66 & 63.06 & \underline{91.23} & \textbf{94.50} & \textit{82.62} \\
18  & 43.52 & 52.39 & 51.56 & 62.42 & 63.55 & 73.94 & 63.11 & \underline{92.69} & \textbf{94.59} & \textit{83.15} \\
20  & 43.94 & 52.26 & 51.50 & 63.06 & 64.68 & 75.07 & 63.44 & \underline{93.59} & \textbf{94.97} & \textit{83.27} \\

\end{tabular}
\label{tab:classification_accuracy}
\end{table*}
\subsection{Performance Benchmark across Different SNR Conditions}

We compared our G-AMC with traditional machine learning algorithms (kNN, DT, and SVM), feature-based algorithm (ML-XGBoost), deep learning methods (VGG, ResNet, CNN-AMC, MCNet), and lightweight deep learning method (lightweight DLAMC). The benchmark is summarized in the TABLE \ref{tab:classification_accuracy}. The G-AMC model outperforms all traditional and feature-based machine learning algorithms. G-AMC can have performance comparable to the deep learning methods. Under the high SNR conditions, it can outperform half of the deep learning methods and rank in the top 3.

\section{Conclusion}
In summary, a mathematically transparent Green Automatic Modulation classification (G-AMC) pipeline was proposed to address the modulation classification problem. The inverse-pyramid architecture, incorporating multi-resolutions from the input signal, was represented by the sparse coding method. Several domain-knowledge features were designed to simplify downstream hierarchical classifiers. Furthermore, hierarchical classifiers were designed to make the best use of the model parameters. The experimental results demonstrated the high computational efficiency and effectiveness of our hierarchical classifier design and integrated features. In future work, we expect to improve the robustness in the low-SNR conditions and the accuracy of high-order modulations.

\bibliographystyle{IEEEtran}   % or another style if you're using a different format
\bibliography{refs}            % if your file is refs.bib
\end{document}